# Cloud Computing Simulation Using CloudSim

Ranjan Kumar[#1], G.Sahoo[*2]

[#]*Assistant Professor, Computer Science & Engineering, Ranchi University, India*
*Professor & Head, Information Technology, Birla Institute of Technology, Mesra, India*

*Abstract*— : As we know that Cloud Computing is a new paradigm in IT. It has many advantages and disadvantages. But in future it will spread in the whole world. Many researches are going on for securing the cloud services. Simulation is the act of imitating or pretending. It is a situation in which a particular set of condition is created artificially in order to study that could exit in reality. We need only a simple Operating System with some memory to startup our Computer. All our resources will be available in the cloud.

## I. INTRODUCTION

Cloud Computing infrastructure is different from Grid Computing. It is the massive deployment of Virtualization technologies and tools. Hence, as compared to Grid, Cloud has an extra layer as Virtualization, that acts as an execution and hosting environment for cloud-based application services. To secure data in cloud is really very tough job. To understand the cloud computing we need to first understand how these resources are placed in the cloud. Cloud Computing has basically two parts, the First part is of Client Side and the second part is of Server Side. The Client Side requests to the Servers and the Server responds to the Clients. The request from the client firstly goes to the Master Processor of the Server Side. The Master Processor have many Slave Processors, the master processor sends that request to any one of the Slave Processor which is free at that time. All Processors are busy in their assigned job and non of the Processor get Idle. Simulation opens the possibility to evaluate the hypothesis prior to actual software development in an environment where one can reproduce tests. Why we need simulation because it provides repeatable and controllable environment to test the services. It tune the system bottlenecks before deploying on real clouds. For simulation we need a special toolkit named CloudSim. It is basically a Library for Simulation of Cloud Computing Scenarios. It has some features such as it support for modeling and simulation of large scale Cloud Computing infrastructure, including data centers on a single physical computing node. It provides basic classes for describing data centers, virtual machines, applications, users, computational resources, and policies.

CloudSim supports VM Scheduling at two levels : First, at the host level where it is possible to specify how much of the overall processing power of each core in a host will be assigned at each VM. And the second , at the VM level, where the VMs assign specific amount of the available processing power to the individual task units that are hosted within its execution engine.

Typical requirements and models of Cloud Computing are Platform (PaaS), Software (SaaS), Infrastructure (IaaS), Service-based application programming interface (API).

It has some Characteristics such as Low Cost Software, Virtualization Service Orientation, Advance Security, Massive Scale, Resilient Computing, Geographic Distribution, Homogeneity, Broad Network Access, Rapid Elasticity, Resource Pooling etc.

There are basically four deployment models, they are Private Cloud, Public Cloud, Community Cloud, and Hybrid Cloud.

Cloud Infrastructure :

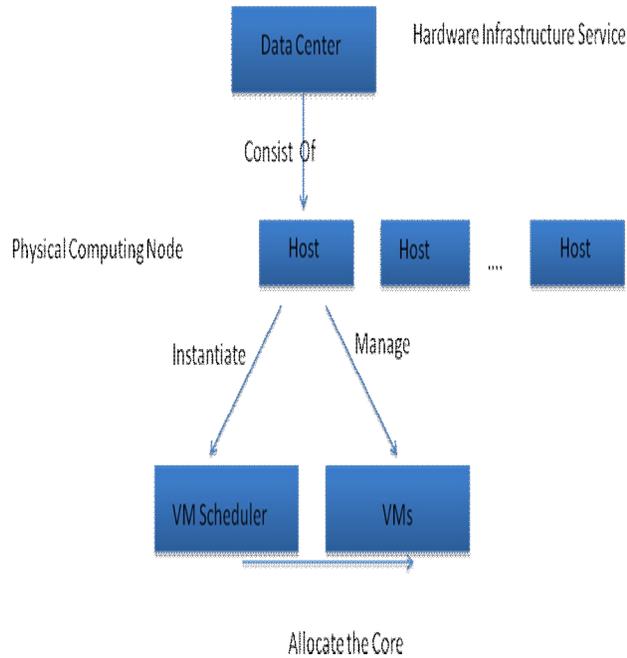

Figure : 1

In figure 1, we saw infrastructure of Cloud, in which Data Center consist of different Hosts and the Host manages the VM Scheduler and VMs.

Cloudlet Scheduler determines how the available CPU resources of virtual machine are divided among Cloudlets. There are two types of policies are offered :

1] Space-Shared (Cloudlet Scheduler Space Shared) : To assign specific CPU cores to specific VMs.

2] Time-Shared (Cloudlet Scheduler Time Shared) : To dynamically distribute the capacity of a core among VMs.

VmSchedular determines how many processing cores of a host are allocated to virtual machines and how many processing cores will be delegated to each VM. It also determine how much of the processing core's capacity will effectively be attributed for a given VM.

In figure 2, we saw the life cycle of CloudSim. In this, there are different stages from where simulation have to pass each of the stage.

CloudSim Life Cycle :

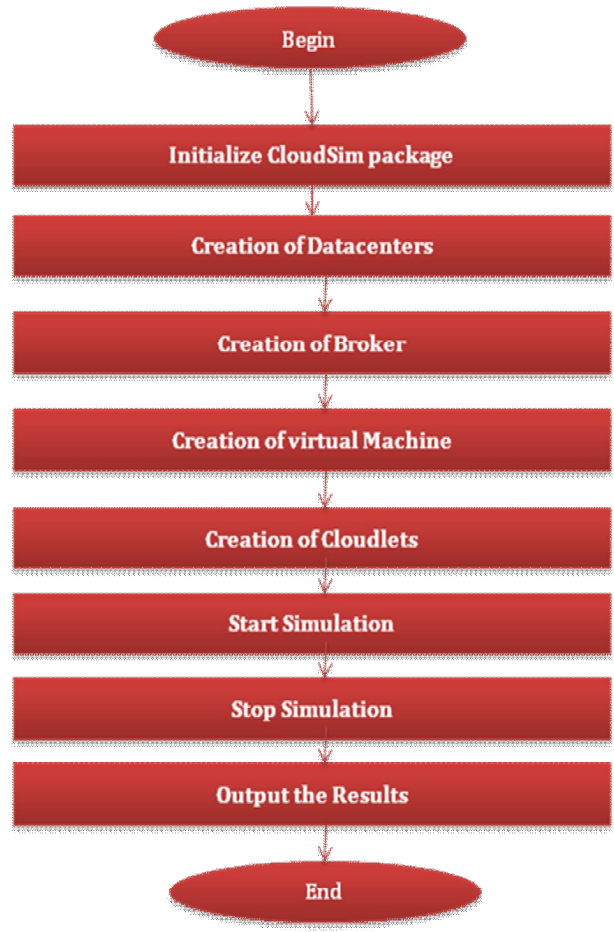

Figure : 2

Organization of this paper is as follows: Related work is discussed in section II. Experimental setup and Result is discussed in section III. And section IV gives conclusion.

## II. RELATED WORK

David S. Linthicum [1] described about the basic information about the Cloud Computing and its various services and models like SaaS, IaaS, PaaS. He also described about the deploying models of Cloud Computing and Virtualization services.

Michael Miller [2] described about the various web based application related to Cloud Computing. R.Bajaj and D.P. Agrawal [3] described about the Scheduling. They said Scheduling is a process of finding the efficient mapping of tasks to the suitable resources so that execution can be completed such as minimization of execution time as specified by customers. They described various types of Scheduling like Static, Dynamic, Centralized, Hierarchical, Distributed, Cooperative, Non-Cooperative Scheduling. They also described Scheduling problem in Cloud and the types of users like CCU ( Cloud Computing Customers) and CCSP ( Cloud Computing Service Providers). R.N. Calheiros, Rajiv Ranjan, Anton Beloglazov, C.A.F. De Rose, Rajkumar Buyya [4] described about the Simulation techniques and the CloudSim. They described the various features of CloudSim like it supports for modelling and simulation for large scale of cloud computing infrastructure including data centers on a single physical computing node. J. Li, M. Qiu, X. Qin [5] described about the optimization criterion that is used when making scheduling decision and represents the goals of the scheduling process. The criterion is expressed by the value of objective function which allows us to measure the quality of computed solution and compare it with different solution. C.H.Hsu and T. L. Chen [6] described about the Quality of Service, that is the ability to provide different jobs and users, or to guarantee a certain level of performance to a job. If the QoS mechanism is supported it allows the user to specify desired performance for their jobs. In system with limited resources the QoS support results in additional cost which is related to the complexity of QoS requests and the efficiency of the scheduler when dealing with them.

### III. EXPERIMENTAL SETUP AND RESULT

To evaluate the performance of Cloud, results were simulated in Window 7 basic (64-bit), i3 Processor, 370 M Processor, 2.40 GHz of speed with memory of 3 GB and the language used is Java. First code of Simulation is tested on one Data Center with one Host and run on one Cloudlet.

Initialize the CloudSim Package :

*Int num_user = 1; //number of cloud users*

*Calendar calendar = Calender.getInstance();*

*Boolean trace_flag = false; //mean trace events*

Initialize the CloudSim Library :

*CloudSim.init(num_user, calendar, trace_flag);*

Create Datacenters : These are the resource providers in CloudSim. We need at least one of them to run a CloudSim simulation.

*Datacenter datacenter0 = createDatacenter("Datacenter_0");*

Create Broker :

*DatacenterBroker broker = createBroker();*

*Int brokerId = broker.getId();*

Create one Virtual machine :

*Vmlist = new ArrayList<Vm>();*

VM description :

*Int vmid = 0;*

*Int mips= 1000;*

*Long size = 10000;  //image size (MB)*

*Int ram = 512;  //vm memory (MB)*

*Long bw = 1000;*

*Int pesNumber = 1; //number of cpu*

*String vmm = "Xen"; // VMM name*

Create VM :

*Vm vm = new Vm(vmid, brokerId, mips, pesNumber, ram, bw, size, vmm, new CloudletSchedulerTimeShared());*

Add the VM to the vmList :

*vmList.add(vm1);*

Submit vm list to the broker :

*Broker.submitVmList(vmList);*

Creation of Cloudlets :

*cloudLetList = new ArrayList<Cloudlet>();*

Cloudlet Properties :

*Int id = 0;*

*PesNumber = 1;*

*long length = 400000;*

*long fileSize = 300;*

*long outputSize = 300;*

*UtilizationModel utilizationModel = new UtilizationModelFull();*

*Cloudlet cloudlet = new Cloudlet(id, length, pesNumber, fileSize, outputSize, utilizationModel, utilizationModel1, utilizationModel1);*

*Cloudlet.setUserId(brokerId);*

*Cloudlet.setVmId(vmid);*

Add the Cloudlets to the list :

*cloudLetList.add(cloudlet1);*

Submit cloudlet list to the broker :

*Broker.submitCloudletList(cloudLetList);*

Start Simulation :

*CloudSim.startSimulation();*

*CloudSim.stopSimulation();*

Final step : Print result when simulation is over

*List<Cloudlet> newList = broker.getCloudletReceivedList();*

*printCloudletList(newList);*

*// Print the debt of each user to each datacenter*

*Datacenter0.printDebts();*

The output of this simulation is :
========== OUTPUT ==========
Cloudlet ID   STATUS   Data center ID   VM ID   Time   Start Time   Finish Time
    0       SUCCESS      2         0     400      0      400
*****PowerDatacenter: Datacenter_0*****
User id        Debt
3              35.6
********************************

After increasing the datacenter with two hosts and run two cloudlets on it. The cloudlets run in VMs with different MIPS requirements. The cloudlets will take different time to complete the execution depending on the requested VM performance.
It prints the following output :

========== OUTPUT ==========
Cloudlet ID   STATUS   Data center ID   VM ID   Time   Start Time   Finish Time
    1       SUCCESS      2         1      80      0       80
    0       SUCCESS      2         0     160      0      160
*****PowerDatacenter: Datacenter_0*****
User id        Debt
3              224.8
********************************

## IV. Conclusions

In this paper, we have proposed the codes for simulation. We see the various outputs in which the information about the Cloudlets, Status, Datacenter ID, Virtual Machine ID, Start Time, Finish Time is given. And by changing the number of Host, Datacenters and Cloudlets, we saw the difference.